\documentclass[prd,nofootinbib,preprint,superscriptaddress]{revtex4}

\usepackage{amsmath, amssymb, amsthm, graphicx, epsfig, fancyhdr,epsfig, slashed, mathrsfs}
\usepackage[T1]{fontenc}
\usepackage{tikzsymbols}
\usepackage{natbib}
\usepackage{float}
\usepackage[thinc]{esdiff}
\usepackage{feynmf}
\usepackage{tikzsymbols}
\usepackage{natbib}
\usepackage{ amssymb }
\usepackage{amsmath}
\usepackage{dsfont}
\usepackage{cancel}
\usepackage[titletoc]{appendix}
\usepackage{float}

\usepackage{tikz,xcolor,hyperref}

\usepackage{slashed}

\definecolor{lime}{HTML}{A6CE39}
\DeclareRobustCommand{\orcidicon}{
	\begin{tikzpicture}
	\draw[lime, fill=lime] (0,0) 
	circle [radius=0.2] 
	node[white] {{\fontfamily{qag}\selectfont \tiny ID}};
	\draw[white, fill=white] (-0.0625,0.095) 
	circle [radius=0.007];
	\end{tikzpicture}
	\hspace{-2mm}
}

\foreach \x in {A, ..., Z}{\expandafter\xdef\csname orcid\x\endcsname{\noexpand\href{https://orcid.org/\csname orcidauthor\x\endcsname}
			{\noexpand\orcidicon}}
}

\usepackage{tikz,xcolor,hyperref}

\definecolor{lime}{HTML}{A6CE39}
\DeclareRobustCommand{\orcidicon}{
	\begin{tikzpicture}
	\draw[lime, fill=lime] (0,0) 
	circle [radius=0.2] 
	node[white] {{\fontfamily{qag}\selectfont \tiny ID}};
	\draw[white, fill=white] (-0.0625,0.095) 
	circle [radius=0.007];
	\end{tikzpicture}
	\hspace{-2mm}
}

\foreach \x in {A, ..., Z}{\expandafter\xdef\csname orcid\x\endcsname{\noexpand\href{https://orcid.org/\csname orcidauthor\x\endcsname}
			{\noexpand\orcidicon}}
}

\newcommand{\be}{\begin{equation}}
\newcommand{\ee}{\end{equation}}
\newcommand{\bea}{\begin{eqnarray}}
\newcommand{\eea}{\end{eqnarray}}

\usepackage{amsmath}
\usepackage{braket}
\usepackage{graphicx}
\usepackage{mathrsfs}
\usepackage{xcolor}
\usepackage{float}
\usepackage{gensymb}
\usepackage{amssymb}
\usepackage{slashed}
\usepackage{latexsym}
\usepackage{breqn}

\newcommand{\ba}{\begin{eqnarray}}
\newcommand{\ea}{\end{eqnarray}}
\newcommand{\bi}{\begin{itemize}}
\newcommand{\ei}{\end{itemize}}

\newcommand{\x}{\star}

\newcommand{\arXiv}[2]{\href{http://arxiv.org/pdf/#1}{{\tt #2/#1}}}
\newcommand{\arXivold}[1]{\href{http://arxiv.org/pdf/#1}{{\tt #1}}}

\begin{document}
\title{Anomalies in String-inspired Non-local Extensions of QED}
\author{Fayez Abu-Ajamieh\orcidA{}}
\email{fayezajamieh@iisc.ac.in}
\affiliation{Centre for High Energy Physics; Indian Institute of Science; Bangalore; India}
\author{Pratik Chattopadhyay\orcidB{}}
\email{pratikpc@gmail.com}
\affiliation{School of Mathematical Sciences; University of Nottingham; Nottingham UK}
\author{Anish Ghoshal\orcidB{}}
\email{anish.ghoshal@fuw.ed.pl}
\affiliation{Institute of Theoretical Physics; Faculty of Physics; University of Warsaw; Warsaw; Poland}
\author{Nobuchika Okada\orcidB{}}
\email{okadan@ua.edu}
\affiliation{
Department of Physics and Astronomy; 
University of Alabama;\\ Tuscaloosa; Alabama 35487; USA}
\begin{abstract}
We investigate anomalies in the class of non-local field theories that have been proposed as an ultraviolet completion of 4-D Quantum Field Theory (QFT) with generalizing
the kinetic energy operators to an infinite series of higher derivatives inspired by string field theory
and ghost-free non-local approaches to quantum gravity.
We explicitly calculate the vector and chiral anomalies in a string-inspired non-local extension of QED. We show that the vector anomaly vanishes as required by gauge-invariance and the Ward identity. On the other hand, although the chiral anomaly vanishes to the leading order with massless fermions, it nonetheless does not vanish with the massive fermions and we calculate it to the leading order in scale of non-locality. 
We also calculate the non-local vector and axial currents explicitly, and present an illustrative example by applying our results to the decay of $\pi^{0} \rightarrow \gamma\gamma$.
\end{abstract}
\maketitle
    
\section{Introduction}\label{sec:intro}
It is a well-known fact that strings, being non-local objects by their nature, are free from ultraviolet (UV) divergences \cite{Witten:1985cc, Kostelecky:1988ta, Kostelecky:1989nt, Freund:1987kt, Freund:1987ck, Brekke:1987ptq, Frampton:1988kr, Tseytlin:1995uq, Seiberg:1999vs, Siegel:2003vt, Biswas:2004qu, Calcagni:2013eua, Calcagni:2014vxa}. This fact inspired many physicists into trying to mimic this good UV behavior by formulating non-local QFTs as extensions of local QFTs, where non-locality is introduced to eliminate any UV divergences that could exist in the local case. The general prescription for transforming local QFTs to non-local ones is to introduce non-locality to the kinetic term via an entire function with infinite derivatives. For instance, in the scalar sector one writes
\begin{equation}\label{eq:NL_Action}
S_{NL} = \int d^{4}x \Big[\frac{1}{2}\phi \mathcal{K}(\Box)(\Box + m^{2})\phi -V(\phi)\Big],
\end{equation}
and the form factor $\mathcal{K}$ has the function of smearing the interaction vertex, such that it becomes spatially finite in size, rather than being point-like, thereby making the interaction non-local. Apart from being an entire function of the $\Box$ operator with infinite derivatives so that no new poles are introduced to the theory; there are no conditions on the form of $\mathcal{K}(\Box)$, and any function that has the required properties is acceptable. However, in order for the UV behavior of loop amplitudes to be finite and avoid divergences, a common choice is to use a simple exponential function
\begin{equation}\label{eq:form_factor}
    \mathcal{K}(\Box) \equiv\exp{\Big(\frac{\Box + m^{2}}{\Lambda^{2}}\Big)},
\end{equation}
where $m$ is the mass of the particle, and $\Lambda$ is the scale of non-locality. With this choice of form factors, it is easy to see that at high energies, loop amplitudes behave like $\sim e^{-\frac{s}{\Lambda^{2}}}$, which is suppressed when $s > \Lambda^{2}$ and is thus free from UV divergences. However, the construction in eqs.~(\ref{eq:NL_Action}) and (\ref{eq:form_factor}) is an ansatz not derived from first principles, and should be treated as an Effective Field Theory (EFT) of yet another UV completion above the scale of non-locality. Furthermore, the form factor in eq. (\ref{eq:form_factor}) will render the theory acausal, albeit at a level suppressed by the scale of non-locality (which should be high). This is the same issue that plagues the Lee-Wick theory \cite{Lee:1969fy, Lee:1970iw}\footnote{Actually, the Lee-Wick theory emerges as the LO expansion of the form factor in eq. (\ref{eq:form_factor})}. Causality violation in such theories was discussed in \cite{Coleman}, and \cite{Alvarez:2009af} described how such causality violation could be measured in colliders. In spite of all of this, non-locality introduced this way can still be used to calculate observables.

The form factor in eq. (\ref{eq:form_factor}) is reminiscent of the star product in noncommutative geometries (see for instance \cite{Witten:1985cc, Seiberg:1999vs, Minwalla:1999px}), which is defined as
\begin{equation}\label{eq:star_product}
(\phi_{1} \star \phi_{2})(x) = e^{\frac{i}{2}\Theta^{\mu\nu}\partial_{\mu}^{y}\partial_{\nu}^{z}}\phi_{1}(y)\phi_{2}(z)\Big|_{x=y=z},
\end{equation}
where the matrix $\Theta^{\mu\nu}$ is a real antisymmetric matrix that defines the algebra of the noncomutative geometry
\begin{equation}\label{eq:NC_geo}
[x^{\mu},x^{\nu}] = i \Theta^{\mu\nu},
\end{equation}

However, there are subtle differences between the two. For example, while the form factor in eq. (\ref{eq:form_factor}) is symmetric and Lorentz-invariant, the star product in eq. (\ref{eq:star_product}) is antisymmetric and is not Lorentz-invariant. Nonetheless, the noncommutative geometry encapsulated in the exponent of the star product regulates UV divergences as does the scale of non-locality, and one can think of the scale of non-locality as an emergent scale from the scale on noncommutativity. An example of how the two can be related can be found in \cite{Minwalla:1999px}.

The first serious step towards constructing a realistic non-local QFT was taken in \cite{Biswas:2014yia}, where the non-local version of the Abelian gauge theory was formulated and the corresponding LHC phenomenology was studied. The formulation of non-local QED makes it possible to investigate the effects of the putative non-locality in this sector, such as the possible enhancement/suppression of scattering processes in colliders, the possible effect on Electroweak Precision Observables (EWPO), and its impact on gauge anomalies. 

Local gauge anomalies were first explained in \cite{Adler:1969gk, Adler:1969er, Bell:1969ts}, and it is now understood that the anomaly associated with the vector current vanishes as a direct result of gauge invariance and the Ward identity, whereas the chiral anomaly associated with the axial current is non-vanishing since the axial current is global and cannot be gauged, implying that it cannot be conserved. The first (and to the best of our knowledge, only) study that attempted at investigating the $U(1)$ gauge anomalies in non-local QED was \cite{Clayton:1993gn}, where the authors utilized a novel formalism dubbed the "Shadow Field Formalism", to show that introducing non-locality does not affect the conservation of the vector current, nor does it remove the chiral anomaly. In the present paper, we attempt at extending a similar treatment to the non-local QED version formulated in \cite{Biswas:2014yia}. In particular, we will try to show that the vector anomaly vanishes and that the Ward identity is respected, and we derive the non-local chiral anomaly and the associated non-local Noether currents. We show that our results through explicit calculation using the non-local QED formulation in \cite{Biswas:2014yia}, agree with the results obtained in \cite{Clayton:1993gn}.

This paper is organized as follows: In Section \ref{sec:review}, we review the non-local QED theory introduced in \cite{Biswas:2014yia}. In Section \ref{sec:anomalies} we explicitly calculate the vector and chiral anomalies in non-local QED and we derive the associated Noether current. We relegate some technical detail to the Appendix, then we compare our results with \cite{Clayton:1993gn} and show that they agree. In Section \ref{sec:decay} we apply our findings to the decay process of $\pi^{0} \rightarrow \gamma\gamma$ and use the result to set an experimental bound on the scale on non-locality, and finally we present our conclusions in Section \ref{sec:conslusion}.

\section{Review of Non-local QED}\label{sec:review}
We begin by providing a quick overview of the non-local extension of QED that was derived in \cite{Biswas:2014yia}. The basic idea behind obtaining the non-local version of QED, is to start with the local version, then introduce the non-locality factor represented by the exponential of an entire function of derivatives, such that the action remains gauge-invariant. With this prescription in mind, the non-local version of QED can be written as
\begin{equation}\label{eq:NL-QED}
    \mathcal{L}_{\text{NL}} = -\frac{1}{4} F_{\mu\nu}e^{\frac{\Box}{\Lambda_{g}}}F^{\mu\nu} + \frac{1}{2}\Big[ i\overline{\Psi}e^{-\frac{\nabla^{2}}{\Lambda_{f}^{2}}}(\slashed{\nabla}+m)\Psi + h.c.\Big],
\end{equation}
where $\nabla_{\mu} = \partial_{\mu} + i e A_{\mu}$, which implies 
\begin{equation}\label{eq:cov_derivative}
    \nabla^{2} = \Box +ie(\partial \cdot A + A \cdot \partial) -e^{2}A^{2}.
\end{equation}
Here, we have accommodated for the fact that the scale of non-locality for the fermions and photon could be different in principle. Notice that while we are using the ordinary derivative in the photon's kinetic term, the covariant derivative has to be used in the fermion sector to keep it gauge-invariant. In calculating the non-local QED anomaly, one only needs the Feynman rules for the fermion propagator and 
the interaction vertices. 
The former is easily extracted to be
\begin{equation}\label{eq:fermion_propagator}
    \Pi_{f} = \frac{i e^{\frac{p^{2}}{\Lambda_{f}^{2}}}(\slashed{p}+m)}{p^{2}-m^{2}+i \epsilon} .
\end{equation}
It is easy to see that in the limit $\Lambda_{f} \rightarrow \infty$ one recovers the standard fermion propagator.
On the other hand, extracting the interaction vertex is more subtle, as special care is needed to include the contribution from the covariant derivative in the exponent. To proceed, we expand the covariant derivative in the non-local factor, then only keep the terms at linear order in $A$. The final result is given by
\begin{equation}\label{eq:ffga_vertex}
    V(k_{1},k_{2}) = -\frac{i e}{2}\Bigg[ (k_{1\mu}\slashed{k}_{2}+k_{2\mu}\slashed{k}_{1})\Bigg(\frac{e^{\frac{k_{1}^{2}}{\Lambda_{f}^{2}}}-e^{\frac{k_{2}^{2}}{\Lambda_{f}^{2}}}}{k_{1}^{2}-k_{2}^{2}} \Bigg) + \Big( e^{\frac{k_{1}^{2}}{\Lambda_{f}^{2}}}+e^{\frac{k_{2}^{2}}{\Lambda_{f}^{2}}}\Big)\gamma_{\mu}\Bigg],
\end{equation}
where $k_{1,2}$ are the momenta of the fermions. In the limit $\Lambda_{f} \rightarrow \infty$ one recovers the local QED. We refer the interested reader to \cite{Biswas:2014yia} for the detailed derivation. 

\section{Anomalies in Non-local QED}\label{sec:anomalies}
In this section, we will explicitly calculate the $U(1)$ vector and axial anomalies in the non-local extension of QED formulated in \cite{Biswas:2014yia}. In our calculation, we follow the method presented in \cite{Bilal:2008qx} based on calculating the triangle diagrams regularized via a Pauli-Villars regulator. However, unlike the case of local QED, no regulator is needed to calculate the loop diagrams in non-local QFTs, as they are already super-renormalizable due to the non-locality form factor.
\begin{figure}[!t]
\centering
\includegraphics[width = 0.8\textwidth]{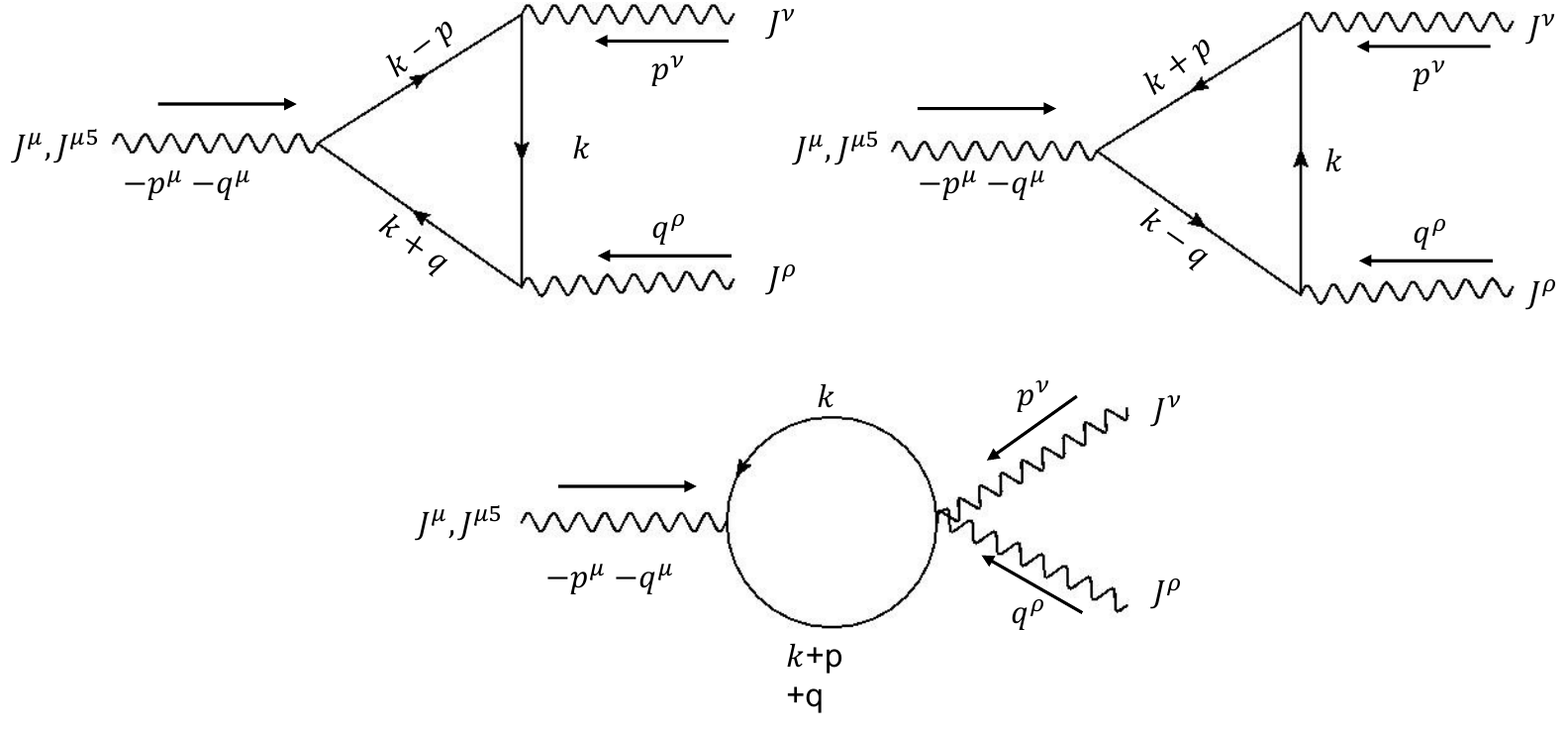}
\caption{\it \small Triangle (top) and bubble (bottom) diagrams contributing to the anomalies in non-local QED.}
\label{fig1}
\end{figure}
Similar to the case of local QFTs, anomalies in non-local QED arise from triangle diagrams with charged fermions running in the loops, with two vector and one axial currents attached to the vertices as shown in the top row of Figure \ref{fig1}. In non-local QED, there is an additional contribution from the bubble diagram shown in the bottom row of Figure \ref{fig1}. One can see how this type of diagrams comes into play by inspecting eqs. (\ref{eq:NL-QED}) and (\ref{eq:cov_derivative}). We can see that when we expand the covariant derivative in the form factor, we obtain an infinite tower of non-renormalizable effective vertices $\sim \overline{\Psi}\Psi A^{n}$, where we see that the bubble diagram arises from the vertex with $n=2$. These interaction vertices are a direct consequence of the requirement of gauge invariance, which necessitated using the covariant derivative instead of the ordinary one in the non-locality form factor. We present the detailed derivation of the Feynman rule associated with the $\overline{\Psi}\Psi A^{2}$ vertex in Appendix \ref{appendix1}.

Before we proceed with calculating the anomalies, we point out that in general, calculating loop diagrams in non-local QFTs is not doable exactly due to the complex nature of the form factor that contains loop momenta to be integrated over. However, the calculation simplifies significantly if we assume that the scale of non-locality is much larger than the external momenta, i.e. $\Lambda \gg p,q$. Given the lower bound on $\Lambda \sim 2.5 - 3$ TeV \cite{Biswas:2014yia}, the validity of this approximation is well-justified, as was demonstrated in detail in \cite{Abu-Ajamieh:2023syy}. In this limit, 
the form factors in the propagators and the interaction vertices are simplified and reduced to $e^{(k\pm p)^{2}/\Lambda^{2}} \simeq e^{(k\pm q)^{2}/\Lambda^{2}} \simeq e^{k^{2}/\Lambda^{2}}$, where $k$ is the loop momentum to be integrated over.

\subsection{Vector and Chiral Anomalies with Massless Fermions}\label{sec:massless}
We first investigate the case where the fermions in the loops are massless. 
We begin by calculating the bubble diagram. In the limit of small external momenta, the corresponding matrix element reads
\begin{eqnarray}\label{eq:Bubble1}
    \mathcal{M}_{\bigcirc}^{\mu\nu\rho} \simeq i e^{2}\int \frac{d^{4}k}{(2\pi)^{4}} e^{\frac{4k^{2}}{\Lambda^{2}}}\text{Tr}\Big[\frac{\gamma^{\mu}\gamma^{5}(\slashed{k}+\slashed{p})V^{\nu\rho}(k+p,k-q,p,q)(\slashed{k}-\slashed{q})}{(k+p)^{2}(k-q)^{2}}\Big],
\end{eqnarray}
where $V^{\nu\rho}$ is given by eq. (\ref{eq:ffAA_Vertex}). Using the explicit expression of $V^{\nu\rho}(k+p,k-q,p,q)$, we find that
\begin{eqnarray}\label{eq:Bubble2}
    \mathcal{M}_{\bigcirc}^{\mu\nu\rho} \sim \text{Tr}\Bigg[\gamma^{\mu}\gamma^{5}(\slashed{k}+\slashed{p})\Big[(\slashed{k}-\slashed{q})(k+p)^{\nu}(k+p-q)^{\rho}-(\slashed{k}+\slashed{p})(k-q)^{\nu}k^{\rho}\Big] (\slashed{k}-\slashed{q})\Bigg] = 0.
\end{eqnarray}
Therefore, the bubble diagram does not contribute to either the vector or the chiral anomalies. On the other hand, the triangle diagrams are given by
\begin{equation}\label{eq:triangle1}
\mathcal{M}_{\triangle}^{\mu\nu\rho} \simeq  - i e^{2} \int \frac{d^{4}k}{(2\pi)^{4}} e^{\frac{6k^{2}}{\Lambda^{2}}} \text{Tr} \Big[ \frac{\gamma^{5}\gamma^{\mu}(\slashed{k}+\slashed{p})\gamma^{\nu}\slashed{k}\gamma^{\rho}(\slashed{k}-\slashed{p})}{(k+p)^{2}k^{2}(k-q)^{2}} \Big] + \begin{pmatrix} p \leftrightarrow q \\ 
\nu \leftrightarrow \rho \end{pmatrix},
\end{equation}
in the limit of small external momenta. Notice that this is identical to the local case multiplied by the non-locality factor. The factor of $6$ arises from 3 non-local vertices and 3 non-local propagators. 

We begin by calculating the vector anomaly. Our aim is to verify that the vector anomaly  indeed vanishes in the non-local QED and that the Ward identity is preserved. Prima facie, this should be the case, since the non-local QED action is gauge invariant by construction. To this avail, it is convenient to calculate $p_{\nu}M_{5}^{\mu\nu\rho}$. Using $\slashed{p} = \slashed{p}+\slashed{k} -\slashed{k}$, the trace in eq. (\ref{eq:triangle1}) simplifies to
\begin{equation}\label{eq:traces1}
    \frac{1}{k^{2}(k-q)^{2}}\text{Tr}\Big[ \gamma^{5}\gamma^{\mu}\slashed{k}\gamma^{\rho}(\slashed{k}-\slashed{q})\Big] - \frac{1}{(k+p)^{2}(k-q)^{2}}\text{Tr}\Big[ \gamma^{5}\gamma^{\mu} (\slashed{k}+\slashed{p})\gamma^{\rho}(\slashed{k}-\slashed{q})\Big].
\end{equation}
It is a simple exercise to evaluate the traces. The first traces yields $-4ik_{\nu}q_{\sigma}\epsilon^{\mu\nu\rho\sigma}$, whereas the second trace evaluates to $-4i(k_{\nu}p_{\sigma}+k_{\nu}q_{\sigma}+p_{\nu}q_{\sigma})\epsilon^{\mu\nu\rho\sigma}$. Thus, eq. (\ref{eq:triangle1}) becomes
\begin{equation}\label{eq:vector_massless1}
p_{\nu}\mathcal{M}_{\triangle}^{\mu\nu\rho} \simeq  - 4 e^{2} \epsilon^{\mu\nu\rho\sigma} \int \frac{d^{4}k}{(2\pi)^{4}} e^{\frac{6k^{2}}{\Lambda^{2}}}  \Bigg[ \frac{k_{\mu}q_{\sigma}}{k^{2}(k-q)^{2}} +\frac{k_{\nu}(p+q)_{\sigma}+ p_{\nu}q_{\sigma}}{(k-q)^{2}(k+p)^{2}} \Bigg] + \begin{pmatrix} p \leftrightarrow q \\ 
\nu \leftrightarrow \rho \end{pmatrix}.
\end{equation}

It is sufficient to evaluate the first term. Focusing on the first part of the first term, we notice that the only external momentum it contains is $q_{\sigma}$, which means that after integrating over $k_{\mu}$, Lorentz invariance implies that the result will be proportional to $q_{\mu}q_{\sigma}$, which vanishes upon contraction with $\epsilon^{\mu\nu\rho\sigma}$. This leaves us with the second integral to perform. Such loop integrals are fairly simple to evaluate and are UV-finite due to the non-locality form factor. Details on how to calculate these non-local momentum integrals are provided in \cite{Abu-Ajamieh:2023syy}. Upon evaluating the momentum integral in eq. (\ref{eq:vector_massless1}), we find 
\begin{equation}\label{eq:vector_massless2}
p_{\nu}\mathcal{M}_{\triangle}^{\mu\nu\rho} \sim (p_{\nu}p_{\sigma} - q_{\nu}q_{\sigma} - p_{\nu}q_{\sigma} - q_{\nu}p_{\sigma})\epsilon^{\mu\nu\rho\sigma}, 
\end{equation}
and we can see that $p_{\nu}p_{\sigma}$ and $q_{\nu}q_{\sigma}$ vanish upon contraction with $\epsilon^{\mu\nu\rho\sigma}$. This leaves $(p_{\nu}q_{\sigma} + q_{\nu}p_{\sigma})\epsilon^{\mu\nu\rho\sigma}$, and it's easy to see that after relabeling $\nu \leftrightarrow \sigma$ in the second term and using the anti-symmetry of $\epsilon^{\mu\nu\rho\sigma}$, the whole term vanishes. The same argument holds for $q_{\nu}\mathcal{M}_{\triangle}^{\mu\nu\rho}$ since $p$ and $q$ are symmetric. Thus, we can see that vector anomaly vanishes in non-local QED, as it should. 

Turning our attention to the chiral anomaly, we need to calculate $(p+q)_{\mu} \mathcal{M}_{\triangle}^{\mu\nu\rho}$. Using
\begin{equation}\label{eq:trick1}
    \gamma^{5}(\slashed{p}+\slashed{q}) = \gamma^{5}(\slashed{p}+\slashed{k} - \slashed{k} +\slashed{q}) = \gamma^{5}(\slashed{k}+\slashed{p}) +(\slashed{k} -\slashed{q})\gamma^{5},
\end{equation}
the trace in (\ref{eq:triangle1}) simplifies to
\begin{equation}\label{eq:traces2}
    \frac{1}{k^{2}(k-q)^{2}}\text{Tr}\Big[ \gamma^{5} \gamma^{\nu}\slashed{k}\gamma^{\rho}(\slashed{k}-\slashed{q})\Big] + \frac{1}{k^{2}(k+p)^{2}}\text{Tr} \Big[ \gamma^{5} (\slashed{k}+ \slashed{p}) \gamma^{\nu} \slashed{k} \gamma^{\rho}\Big].
\end{equation}
Notice that the first term is identical to the first term in (\ref{eq:traces1}) with $\mu \rightarrow \nu$, and therefore it vanishes as we saw above. On the other hand, the second traces yields $-4i\epsilon^{\mu\nu\rho\sigma}k_{\mu}p_{\sigma}$. Therefore, the chiral anomaly reads
\begin{equation}\label{eq:chiral_massless1}
    -(p+q)_{\mu}\mathcal{M}_{\triangle}^{\mu\nu\rho} \simeq 4 e^{2} \epsilon^{\mu\nu\rho\sigma}  \int \frac{d^{4}k}{(2\pi)^{4}} e^{\frac{6k^{2}}{\Lambda^{2}}}\Bigg[\frac{k_{\mu}p_{\sigma}}{k^{2}(k+p)^{2}} \Bigg] + \begin{pmatrix} p \leftrightarrow q \\ 
\nu \leftrightarrow \rho \end{pmatrix}, 
\end{equation}
and we see that the first term contains $p$ only, which means that after integrating over $k$, the result will be $\sim p_{\mu}p_{\sigma}$, which vanishes upon contraction with $\epsilon^{\mu\nu\rho\sigma}$, i.e. the chiral anomaly seems to vanish in non-local QED! This result is counter-intuitive, as the chiral anomaly in local QED is non-vanishing, and one would expect the same to carry on to the non-local case. The reason behind this apparent contradiction lies in the our approximations. We limited our calculation to the leading-order in the expansion of $p,q/\Lambda$, and assumed massless fermions. However, this situation does not hold once we include the NLO expansion in external momenta and/or we use massive fermions, and the chiral anomaly no longer vanishes. In \ref{sec:massive} below, we shall redo our calculation with massive fermions and show that chiral anomaly indeed persist. We will limit our calculation to the LO in $p,q/\Lambda$ for simplicity. 

\subsection{Vector and Chiral Anomalies with Massive Fermions}\label{sec:massive}
Here we show the effect of including fermion masses on both the vector and chiral anomalies. First, let us focus on the bubble diagram. Including the fermion masses in eq. (\ref{eq:Bubble1}) simplifying it, 
eq.~(\ref{eq:Bubble2}) becomes
\begin{eqnarray}\label{eq:Bubble3}
    \mathcal{M}_{\bigcirc}^{\mu\nu\rho} \simeq i e^{2}\int \frac{d^{4}k}{(2\pi)^{4}} e^{\frac{4k^{2}}{\Lambda^{2}}}\text{Tr}\Big[\frac{\gamma^{\mu}\gamma^{5}(\slashed{k}+\slashed{p}+m)V^{\nu\rho}(k+p,k-q,p,q)(\slashed{k}-\slashed{q}+m)}{[(k+p)^{2}-m^{2}][(k-q)^{2}-m^{2}]}\Big],
\end{eqnarray}
with $V^{\nu\rho}(k+p,k-q,p,q)$ which is unchanged compared to the massless case. 
Here too, we find that the trace vanishes, and hence the bubble diagram does not contribute. 
On the other hand, the contribution of the triangle diagrams in eq. (\ref{eq:triangle1}) becomes 
\begin{equation}\label{eq:triangle2}
\mathcal{M}_{\triangle}^{\mu\nu\rho} \simeq  - i e^{2} \int \frac{d^{4}k}{(2\pi)^{4}} e^{\frac{6k^{2}}{\Lambda^{2}}} \text{Tr} \Bigg[ \frac{\gamma^{5}\gamma^{\mu}(\slashed{k}+\slashed{p}+m)\gamma^{\nu}(\slashed{k}+m)\gamma^{\rho}(\slashed{k}-\slashed{p}+m)}{[(k+p)^{2}-m^{2}][k^{2}-m^{2}][(k-q)^{2}-m^{2}]} \Bigg] + \begin{pmatrix} p \leftrightarrow q \\ 
\nu \leftrightarrow \rho \end{pmatrix}.
\end{equation}

First, we investigate the vector anomaly by calculating $p_{\nu}\mathcal{M}_{\triangle}^{\mu\nu\rho}$. Simplifying the expression by writing $\slashed{p} = (\slashed{p}+\slashed{k}-m)-(\slashed{k}-m)$ and then evaluating the traces explicitly, it's not hard to see that the result is identical to eq. (\ref{eq:vector_massless1}) with the denominators being those of massive fermions. Therefore, the result in eq. (\ref{eq:vector_massless2}) continues to hold, and the vanishing of the vector anomaly remains unaffected, as is expected.

Turning our attention to the chiral anomaly by considering $-(p+q)_{\mu}\mathcal{M}_{\triangle}^{\mu\nu\rho}$ in the massive case, we first simplify the matrix element by using
\begin{eqnarray}\label{eq:trick2}
     \gamma^{5}(\slashed{p} + \slashed{q}) & = & \gamma^{5}(\slashed{k}+\slashed{p} - m) + \gamma^{5}(\slashed{q} - \slashed{k} -m) + 2m \gamma^{5} \nonumber \\
     & = & \gamma^{5}(\slashed{k}+\slashed{p} - m) + (\slashed{k} -\slashed{q} -m)\gamma^{5} + 2m \gamma^{5},
\end{eqnarray}
which simplifies the trace in eq.~(\ref{eq:triangle2}) to
\begin{eqnarray}\label{eq:traces3}
      & = & \text{Tr}\Big[\frac{ \gamma^{5}\gamma^{\nu}(\slashed{k}+m)\gamma^{\rho}(\slashed{k}-\slashed{q}+m)}{[k^{2}-m^{2}][(k-q)^{2}-m^{2}]}\Big] + \text{Tr}\Big[\frac{ \gamma^{5}(\slashed{k}+\slashed{p}+m)\gamma^{\nu}(\slashed{k}+m)\gamma^{\rho}}{[(k+p)^{2}-m^{2}][k^{2}-m^{2}]}\Big] \nonumber \\
      & + & 2 m \text{Tr}\Big[ \frac{\gamma^{5}(\slashed{k}+\slashed{p}+m)\gamma^{\nu}(\slashed{k}+m)\gamma^{\rho}(\slashed{k}-\slashed{q}+m)}{[(k-q)^{2}-m^{2}][(k+p)^{2}-m^{2}][k^{2}-m^{2}]}\Big].
\end{eqnarray}

Focusing first and second terms, it is a simple exercise to show that they yield identical results to eq. (\ref{eq:traces2}) with the mass added in the denominators, and therefore they vanish after integrating over $k$ and contracting with $\epsilon^{\mu\nu\rho\sigma}$. The last term, on the other hand, is proportional to the mass and does not yield a vanishing contribution. The trace yields the factor $4i m p_{\mu}q_{\sigma}\epsilon^{\mu\nu\rho\sigma}$, and thus eq. (\ref{eq:triangle2}) becomes
\begin{equation}\label{eq:chiral_massive1}
    -(p+q)_{\mu}\mathcal{M}_{\triangle}^{\mu\nu\rho} \simeq \int \frac{d^{4}k}{(2\pi)^{4}} e^{\frac{6k^{2}}{\Lambda^{2}}} \frac{-8 e^{2} m^{2}p_{\mu}q_{\sigma}\epsilon^{\mu\nu\rho\sigma}}{[(k-q)^{2}-m^{2}][(k+p)^{2}-m^{2}][k^{2}-m^{2}]} + \begin{pmatrix} p \leftrightarrow q \\ 
\nu \leftrightarrow \rho \end{pmatrix}.
\end{equation}

Evaluating the integral is fairly straightforward, and the result in terms of the Feynman parameters reads
\begin{equation}\label{eq:chiral_massive2}
    -(p+q)_{\mu}\mathcal{M}_{\triangle}^{\mu\nu\rho} \simeq \frac{ie^{2}}{\pi^{2}}p_{\mu} q_{\sigma} \epsilon^{\mu\nu\rho\sigma}\int_{0}^{1}dxdy \Big[\frac{1}{1-xy \frac{Q^{2}}{m^{2}}} + \frac{6m^{2}}{\Lambda^{2}}+\frac{12m^{2}}{\Lambda^{2}}\text{Ei}\Big( \frac{6(x y Q^{2}-m^{2})}{\Lambda^{2}}\Big)\Big],
\end{equation}
where $Q^{2} \equiv (p+q)^{2}$, and the exponential integral function $\text{Ei}(x)$ is defined as
\begin{equation}\label{eq:Ei(x)}
    \text{Ei}(x) = - \int_{-x}^{\infty}dt \frac{e^{-t}}{t}.
\end{equation}

Linking eq. (\ref{eq:chiral_massive2}) to the massless case is straightforward and can be done simply by taking the limit $m \rightarrow 0$, which leads to the vanishing of the anomaly at LO in the expansion of the external momenta, in a manner consistent with what we found in Section \ref{sec:massless}. On the other hand, the link to the local case is more subtle. Here one expects that the local case should be obtained by taking the limit $\Lambda \rightarrow \infty$, however, this turns out to be insufficient. The reason behind this can be best understood by calculating the local anomaly following the method in \cite{Bilal:2008qx}, where it is shown that the chiral anomaly in the local case arises purely from the regulator. However, a regulator is absent in the non-local case since it's already finite. Therefore, simply taking $\Lambda \rightarrow \infty$ will not render the regularized local result. Instead, we use the following prescription to remedy the situation: We assume that $m^{2} \gg Q^{2}$, which \textit{corresponds to the mass itself acting as regulator}. In the limit $\Lambda \gg m^{2} \gg Q^{2}$, eq. (\ref{eq:chiral_massive2}) becomes 
\begin{equation}\label{eq:chiral_massive3}
    -(p+q)_{\mu}\mathcal{M}_{\triangle}^{\mu\nu\rho} \simeq \frac{ie^{2}}{2\pi^{2}}p_{\mu} q_{\sigma} \epsilon^{\mu\nu\rho\sigma}\Big[1 + \frac{6m^{2}}{\Lambda^{2}}+\frac{12m^{2}}{\Lambda^{2}}\text{Ei}\Big( \frac{-6m^{2}}{\Lambda^{2}}\Big)\Big],
\end{equation}
and it's easy to see that upon taking $\Lambda \rightarrow \infty$, the local case is retrieved.

\subsection{Noether Currents}\label{sec:Noether}
Finally, here we derive the non-local Noether vector and axial currents. Notice that the action in eq. (\ref{eq:NL-QED}) is invariant under the global transformations
\begin{equation}\label{eq:Global_transformations}
    \Psi \rightarrow e^{i\alpha}\Psi, \hspace{10mm} \Psi \rightarrow e^{i\beta \gamma^{5}}\Psi.
\end{equation}
To derive the corresponding Noether currents, we follow the usual prescription of demanding that the Lagrangian be invariant under the infinitesimal local transformations
\begin{equation}\label{eq:local_transformations}
    \Psi \rightarrow (1+i\alpha(x))\Psi, \hspace{10mm} \Psi \rightarrow (1+ i\beta(x) \gamma^{5})\Psi,
\end{equation}
which leads to the current
\begin{equation}\label{eq:Noether1}
    J^{\mu}(x) = \frac{\delta \mathcal{L}}{\delta(\partial_{\mu}\Psi)}\Delta \Psi.
\end{equation}

In order to derive the non-local QED Noether currents, we start with the Lagrangian
\begin{equation}\label{eq:NL_Noether_Lag0}
    \mathcal{L} = \frac{i}{2}\overline{\Psi}\exp{\Bigg( \frac{-\Box -ie(\partial \cdot A +A \cdot \partial) -e^{2}A^{2}}{\Lambda^{2}}\Bigg)}(\slashed{\partial}\Psi + i e \slashed{A}\Psi) + \text{h.c.}
\end{equation}
Notice that in order to evaluate the variation of the Lagrangian w.r.t. $\partial \Psi$, we need to pay special attention to the derivatives in the exponent. To this avail, we use the following prescription: First we expand the derivative operators in the exponents, then we act the derivatives on the associated field leaving only terms $\sim \partial \Psi$. Finally we exponentiate the results and restore the operator form in the currents. Let us first focus on the second term in the parentheses in eq.~(\ref{eq:NL_Noether_Lag0}). 
We assume that the photon is on-shell, such that $\Box (\slashed{A}\Psi) = \slashed{A}\Box \Psi = -k_{1}^{2} \slashed{A} \Psi$. Therefore we have
\begin{equation}\label{eq:Noether1_1}
    \exp{\Big( -\frac{\Box}{\Lambda^{2}}\Big)}(\slashed{A}\Psi) = \sum_{n=0}^{\infty}\Big[ \frac{(-i)^{n}\Box^{n}}{\Lambda^{2n}n!} \Big] (\slashed{A}\Psi) = \sum_{n=0}^{\infty}\Big[ \frac{(k_{1}^{2})^{n}}{\Lambda^{2n}n!} \Big] (\slashed{A}\Psi) =  \exp{\Big( \frac{k_{1}^{2}}{\Lambda^{2}}\Big)}(\slashed{A}\Psi).
\end{equation}

On the other hand, the remaining derivative acting on $\slashed{A}\Psi$ can be evaluated as follows: 
\begin{eqnarray}\label{eq:Noether1_2}
    \exp{\Big( \frac{-i e A\cdot \partial}{\Lambda^{2}}\Big)} (i e \slashed{A} \Psi) & = & i e \sum_{n=0}^{\infty} \frac{(-i e A \cdot \partial)^{n}}{\Lambda^{2n}n!} (\slashed{A}\Psi), \nonumber \\
    & = &  i e \sum_{n=0}^{\infty} \frac{(-i e A^{\mu})^{n}}{\Lambda^{2n}n!} \sum_{k=0}^{n} {n \choose k} (\partial_{\mu}^{n-k}\slashed{A})(\partial_{\mu}^{k}\Psi), \nonumber \\
    & = &  i e \slashed{A} A \cdot \partial \Psi \sum_{n=0}^{\infty} \frac{(-i e A^{\mu})^{n}}{\Lambda^{2n}n!} \sum_{k=0}^{n} {n \choose k} (i q \cdot A)^{n-k} (-i k_{1} \cdot A)^{k-1}, \nonumber \\
    & = & - \frac{e^{2}\slashed{A}A\cdot \partial \Psi}{k_{1}\cdot A} \exp{\Big( \frac{e A \cdot k_{2}}{\Lambda^{2}}\Big)},
\end{eqnarray}
where we have used conservation on momentum to eliminate the momentum of the photon. The hermitian conjugate yields identical results with $k_{1} \leftrightarrow k_{2}$. Thus, after restoring the operators, the second term in eq. (\ref{eq:NL_Noether_Lag0}) becomes
\begin{equation}\label{eq:Noether1_3}
   \mathcal{L}_{2} = -\frac{i e^{2}}{2} \overline{\Psi} \exp{\Bigg( \frac{-\Box -i e A \cdot \partial - e^{2}A^{2}}{\Lambda^{2}}\Bigg)} \Big(\frac{1}{k_{1}\cdot A} + \frac{1}{k_{2}\cdot A} \Big) \slashed{A}A \cdot \partial \Psi. 
\end{equation}
Notice that when the photon is assumed to be on-shell, we have
\begin{eqnarray}\label{eq:vanishing}
    \frac{1}{k_{1}\cdot A} + \frac{1}{k_{2}\cdot A} = \frac{(k_{1}+k_{2})\cdot A}{(k_{1}\cdot A)(k_{2}\cdot A)} = \frac{q \cdot A}{(k_{1}\cdot A)(k_{2}\cdot A)} = 0,
\end{eqnarray}
which implies that the second term in eq. (\ref{eq:NL_Noether_Lag0}) does not contribute to the non-local Noether currents. On the other hand, the first term will give a non-vanishing contribution. Follwing the same procedure, we obtain
\begin{equation}\label{eq:NL_Noether_Lag1}
    \mathcal{L}_{1} = \frac{i}{2}\overline{\Psi}\exp{\Bigg( \frac{k_{1}^{2} + e A \cdot k_{2} -e^{2}A^{2}}{\Lambda^{2}}\Bigg)} \slashed{\partial}\Psi +(1 \leftrightarrow 2).
\end{equation}
Using eq. (\ref{eq:NL_Noether_Lag1}) in eq. (\ref{eq:Noether1}), then restoring the operators in the exponents, we obtain the Noether currents 
\begin{eqnarray}
    J^{\mu}(x) & = &  \overline{\Psi} \gamma^{\mu} \Psi \exp{\Big(\frac{-\Box - i e A \cdot \partial -e^{2}A^{2}}{\Lambda^{2}} \Big)} , \label{eq:J1}\\
    J^{\mu 5}(x) & = & \overline{\Psi} \gamma^{\mu} \gamma^{5} \Psi \exp{\Big(\frac{-\Box - i e A \cdot \partial -e^{2}A^{2}}{\Lambda^{2}} \Big)}. \label{eq:J2} 
\end{eqnarray}
Notice that taking the limit $\Lambda \rightarrow \infty$, the local limit is retrieved, i.e $J^{\mu} \rightarrow \overline{\Psi}\gamma^{\mu}\Psi$, and $J^{\mu 5} \rightarrow \overline{\Psi}\gamma^{\mu} \gamma^{5}\Psi$.

Before we conclude this section, there is an important point that we need to clarify. As is well-known, local anomalies are obtained by evaluating the expectation of the Noether currents. Thus, we should be able to obtain the non-local anomalies by evaluating 
\begin{equation}\label{eq:expectation}
 \int d^{4}x d^{4}y d^{4}z e^{-ip.x} e^{iq_{1}.y} e^{iq_{2}.z}\langle J^{\mu 5}(x)J^{\nu}(y)J^{\rho}(z) \rangle,
\end{equation}
with the currents given by eqs. (\ref{eq:J1}) and (\ref{eq:J2}). However, given the field $A$ in the exponents of the vector and axial currents, we see that the expansion in $A$ actually corresponds to the sum of all insertions of the vector current in the fermion loop, i.e. the quantity in eq. (\ref{eq:expectation}) actually encodes all higher-order anomalies that correspond to an arbitrary number of the gauge field $A$ inserted into a fermion loop (in addition to the insertions of vector and axial fields from the local piece). These anomalies in general, might not be vanishing, however, we are only interested in the triangle anomalies. Triangle anomalies can be obtained by keeping the leading order in $A$, i.e.
\begin{equation}\label{eq:A_LO}
    \exp{\Big(\frac{-\Box - i e A \cdot \partial -e^{2}A^{2}}{\Lambda^{2}} \Big)} \simeq \exp{\Big(-\frac{\Box}{\Lambda^2} \Big)} + O(A).
\end{equation}
Thus we can see at this order, that eq.~(\ref{eq:expectation}) leads to the same results we obtained above.

\subsection{Summary of the Results}\label{sec:Summaru}
In this section we summarize the results that we obtained in this paper: 

\begin{itemize}  

\item Vector anomalies in non-local QED vanish exactly, whether the fermions in the loops are massless or massive, and the Ward identity is respected. It is also not hard to show that the vanishing of the vector anomaly holds to all orders in the expansion of $p,q/\Lambda$. This is expected, since the non-local QED action in eq. (\ref{eq:NL-QED}) is gauge-invariant by construction, 

\item Although in non-local QED with massless fermions, the chiral anomaly appears to vanish at the LO in $p,q/\Lambda$; one can show that is no longer holds once higher-order corrections are included. In addition, for non-local QED with massive fermions at LO, we find that the chiral anomaly persists and that it has the expected form. We found that while obtaining the massless limit is straightforward, the local limit is more subtle and cannot be obtained by simply taking $\Lambda \rightarrow \infty$. Instead, one needs to assume that mass of the fermions is much larger than the other momentum scales in order to act as a regulator itself in the local limit. Using this prescription, the correct local limit is obtained,

\item The non-local vector and axial currents encode anomalies that correspond to all insertions of the gauge field in the fermion loop, with the triangle anomalies obtained from the LO expansion in the gauge field. This is a direct consequence of gauge invariance, which leads to rich structures in non-local QED that merit further investigation in the future. 

\item Our results are consistent with those found in Ref. \cite{Clayton:1993gn} using the shadow field formalism.
\end{itemize}

\section{Application: $\pi^{0} \rightarrow \gamma\gamma$ Decay}\label{sec:decay}
We present an application to anomalies in non-local QED by studying the decay process of $\pi^{0} \rightarrow \gamma\gamma$. This decay proceeds through triangle diagrams like the ones shown in Figure \ref{fig1}, with the axial current replaced with a pseudo-scalar and with protons running in the loops. The interaction Lagrangian is given by
\begin{equation}\label{eq:interaction}
    \mathcal{L}_{\text{int}} = - i\lambda \pi \overline{\Psi}\gamma^{5}\Psi.
\end{equation}

The matrix element can be written as $-\lambda e^{2}\epsilon_{1\mu}^{*}\epsilon_{2\nu}^{*}\mathcal{M}^{\mu\nu}$, where at LO in $q_{1,2}/\Lambda$ we have

\begin{equation}\label{eq:matrix2}
    \mathcal{M}^{\mu\nu} \simeq \int \frac{d^{4}k}{(2\pi)^{4}} e^{\frac{5k^{2}}{\Lambda^{2}}}\text{Tr}\Bigg[ \gamma^{\mu} \frac{i(\slashed{k}-\slashed{q_{1}}+m)}{(k-q_{1})^{2}-m^{2}}\gamma^{5}\frac{i(\slashed{k}+\slashed{q}_{2}+m)}{(k+q_{2})^{2}-m^{2}}\gamma^{\nu}\frac{i(\slashed{k}+m)}{k^{2}-m^{2}} \Bigg] + \begin{pmatrix} 1 \leftrightarrow 2 \\ 
\nu \leftrightarrow \rho \end{pmatrix},
\end{equation}
where $m$ is the mass of the proton. $\mathcal{M}^{\mu\nu}$ can be evaluated following the procedure illustrated in Section \ref{sec:anomalies}, and in the limit $m \gg m_{\pi}$, the decay width reads
\begin{equation}\label{eq:NL_decay_wdith}
    \Gamma_{\text{NL}}(\pi^{0}\rightarrow\gamma\gamma) \simeq \Gamma_{0} \times \Bigg[1 + \frac{5m^{2}}{\Lambda^{2}} + \frac{10m^{2}}{\Lambda^{2}} \text{Ei}\Big( -\frac{5m^{2}}{\Lambda^{2}}\Big) \Bigg]^{2},
\end{equation}
where 
\begin{equation}\label{eq:L_decay_wdith}
     \Gamma_{0} = \frac{\alpha^{2}}{64\pi^{3}} \frac{m_{\pi}^{3}}{f_{\pi}^{2}},
\end{equation}
is the decay width in the local case, and $f_{\pi}$ is the pion decay constant. We can use eq. (\ref{eq:NL_decay_wdith}) to set a lower limit on the scale of non-locality. The most recent measurement of the decay width of $\pi^{0} \rightarrow \gamma\gamma$ comes from the PrimEx-II experiment: 
\begin{equation}\label{eq:measurment}
     \Gamma_{\text{Exp}}(\pi^{0}\rightarrow\gamma\gamma) = 7.802 \pm 0.052 \hspace{1mm} (\text{stat.}) \pm 0.105 \hspace{1mm} (\text{syst.}) \hspace{1mm} \text{eV},
\end{equation}
which can be used to set a $2\sigma$ limit on the scale of non-locality
\begin{equation}\label{eq:bound}
    \Lambda \gtrsim 57 \hspace{1mm} \text{GeV}.
\end{equation}
This bound is not very stringent and cannot compete with the collider bound of $ \Lambda \gtrsim 2.5 - 3$ TeV \cite{Biswas:2014yia}.

\section{Conclusion and Outlook}\label{sec:conslusion}
In this paper, we investigated the vector and chiral anomalies in the non-local QED formulated in \cite{Biswas:2014yia}. We found that the vanishing of the vector anomaly remains unaffected and that the Ward identity continues to hold in the non-local case as well. This is to be expected since non-local QED is gauge invariant by construction.

We also found that at leading order the chiral anomaly vanishes in the massless case, while it does not vanish in the massive case. Also, the anomaly continues to exist at next to leading order in the massless case. Naively, one might speculate that since non-local QED lacks a regulator as it is already regularized, and that since the chiral anomaly in the local case arises purely from the regulator; the chiral anomaly in the non-local case would vanish. Nonetheless, this turned out not to be the case, and the chiral anomaly is none-vanishing at next to leading order for the massless case, and can be expressed in terms of the local anomaly plus corrections suppressed by the scale of non-locality. We found that obtaining the local limit from the non-local case would require special care and we found that with the correct prescription, the local limit is obtained when $\Lambda \rightarrow \infty$. Our results are consistent with the results found in \cite{Clayton:1993gn} by using the shadow field formalism. We also found the corresponding vector and axial Noether currents in the non-local case and found that they encode all higher-order anomalies, with the triangle anomalies obtained from the LO expansion in the gauge field. We also showed that in the limit $\Lambda \rightarrow \infty$, the local currents are obtained.

As a simple application of our results, we calculated the corrections to the decay width $\pi^{0} \rightarrow \gamma\gamma$ due to non-locality and found that constraint corresponding to the current experimental measurement is weak compared to the limit obtained from the LHC.

\section*{Acknowledgment}
FA thanks Sudhir Vempati. The work of FA is supported by the C.V. Raman fellowship from CHEP at IISc. The work of PC is supported by an EPSRC fellowship.
The work of NO is supported in part by the United States Department of Energy Grant, No.~DE-SC0012447.

\appendix
\section{Derivation of the Non-local $\overline{\Psi}\Psi \gamma\gamma$ Vertex}\label{appendix1}
Here we show how to derive the Feynman rule for the $\overline{\Psi}\Psi\gamma\gamma $ vertex in non-local QED. The Feynman rule for $\overline{\Psi}\Psi\gamma$ vertex was derived in \cite{Biswas:2014yia} and is shown
in eq.~(\ref{eq:ffga_vertex}). The full Feynman rule of the the $\overline{\Psi}\Psi\gamma\gamma $ vertex is rather complex, therefore, we simplify by assuming that the photons are $\textit{on-shell}$, which is the case we are interested in for calculating the anomalies, and we only keep the leading terms in $1/\Lambda^{2}$. We start with the fermion part of the non-local QED action in eq.~(\ref{eq:NL-QED})
\begin{equation}\label{eq:NL_action1}
    S_{\text{NL}} = \frac{1}{2}\int d^{4}x \Big[i\overline{\Psi}e^{-\frac{\nabla^{2}}{\Lambda^{2}}} (\slashed{\nabla} + m) \Psi +h.c.\Big].
\end{equation}

We first expand the non-local form factor in powers of $1/\Lambda^{2}$ and write the covariant derivative explicitly as shown in eq.~(\ref{eq:cov_derivative}):
\begin{equation}\label{eq:NL_action2}
     S_{\text{NL}} = \frac{1}{2}\int d^{4}x \Bigg\{i\overline{\Psi} \sum_{n=1}^{\infty}\frac{(-1)^{n}}{\Lambda^{2n}n!}\Big[ \Box + ie(\partial \cdot A + A \cdot \partial)-e^{2}A^{2}\Big]^{n}\Big[\slashed{\partial} + i e \slashed{A}\Big]\Psi + h.c. \Bigg\}.
\end{equation}
In order to obtain the $\overline{\Psi}\Psi\gamma\gamma$ vertex, we only keep terms that are proportional to $A^{2}$, i.e.~the terms $\sim O(e^{2})$. Inspecting eq.~(\ref{eq:NL_action2}), we can see that we can obtain terms at $O(A^{2})$ through 3 different ways: 1) For $n=1$, we can have the $A^{2}$ term in the first bracket multiplied by the $\slashed{\partial} \Psi$ term in the second bracket, 2) for $n=1$, we can have the $(\partial.A + A.\partial)$ term from the first bracket multiplied by the $\slashed{A}$ term in second bracket, and 3) for $n=2$, we can have the $(\partial \cdot A + A \cdot \partial)^{2}$ term from the first bracket multiplied by the $\slashed{\partial} \Psi$ term in the second term. Explicitly, we have
\begin{eqnarray}\label{eq:NL_action3}
    S_{\text{NL}} \supset & -\frac{i e^{2}}{2}\sum\limits_{n=0}^{\infty}\frac{(-1)^{n}}{\Lambda^{2n}n!}\int d^{4}x \Big\{ \sum\limits_{m=0}^{n-1}(\Box^{m}\overline{\Psi})\Big[A^{2}\Box^{n-m-1}(\slashed{\partial}\Psi) + (\partial \cdot A + A \cdot \partial)\Box^{n-m-1}(\slashed{A}\Psi)\Big] \nonumber\\ 
    & + \sum\limits_{m=0}^{n-2}\sum\limits_{l=0}^{n-m-2}  (\Box^{m}\overline{\Psi})(\partial \cdot A + A \cdot \partial)\Box^{l}(\partial \cdot A + A \cdot \partial)\Box^{n-m-l-2}(\slashed{\partial}\Psi) + h.c.\Big\},
\end{eqnarray}
where we have integrated $\overline{\Psi} \Box^{m}$ by parts to obtain $\Box^{m}\overline{\Psi}$. 
We treat each of the three terms separately. Starting with the first term, notice that each $\Box$ operator will pull down a factor of $-k_{1,2}^{2}$, with $k_{1,2}$ being the 4-momentum of $\overline{\Psi}$ and $\Psi$, respectively. On the other hand, the $\slashed{\partial}\Psi$ will pull a factor of $-i \slashed{k}_{2}$, whereas the hermitian conjugate will give a factor of $-i \slashed{k}_{1}$, thereby symmetrizing the result between $k_{1}$ and $k_{2}$. Thus, the first term yields
\begin{equation}\label{eq:1stContribution1}
    S_{1} = \frac{e^{2}}{2}\sum_{n=0}^{\infty}\frac{1}{\Lambda^{2n}n!}\sum_{m=0}^{n-1}\int d^{4}x(\slashed{k}_{1}+\slashed{k}_{2})(k_{1}^{2m}k_{2}^{2(n-m-1)})\overline{\Psi}\Psi A^{2},
\end{equation}
and the sums can be evaluated as follows
\begin{equation}\label{eq:1stContribution_sum1}
    \sum_{n=0}^{\infty}\frac{1}{\Lambda^{2n}n!}\sum_{m=0}^{n-1}k_{1}^{2m}k_{2}^{2(n-m-1)} = \sum_{n=0}^{\infty}\frac{k_{2}^{2n-2}}{\Lambda^{2n}n!} \Bigg[ \frac{1-(k_{1}^{2}/k_{2}^{2})^{n}}{1-(k_{1}^{2}/k_{2}^{2})}\Bigg] = \frac{e^{\frac{k_{2}^{2}}{\Lambda^{2}}}-e^{\frac{k_{1}^{2}}{\Lambda^{2}}}}{k_{2}^{2}-k_{1}^{2}},
\end{equation}
which, together with eq. (\ref{eq:1stContribution1}), implies that the contribution of the first term is given by
\begin{equation}\label{eq:1stContribution2}
    V_{1\mu\nu}(k_{1},k_{2},q_{1},q_{2}) = i e^{2}(\slashed{k}_{1}+\slashed{k}_{2})\Bigg( \frac{e^{\frac{k_{2}^{2}}{\Lambda^{2}}}-e^{\frac{k_{1}^{2}}{\Lambda^{2}}}}{k_{2}^{2}-k_{1}^{2}}\Bigg)g_{\mu\nu},
\end{equation}
where $q_{1,2}$ are the momenta of the photons, which will be relevant for the remaining contributions. Turning to the second term in eq. (\ref{eq:NL_action3}), we have 
\begin{equation}\label{eq:2ndContribution1}
    S_{\text{NL},2} = \frac{ie^{2}}{2} \sum_{n=0}^{\infty}\frac{1}{\Lambda^{2n}n!}\sum_{m=0}^{n-1} \int d^{4}x \Big[ (k_{1}^{2m}k_{2}^{2(n-m-1)})\overline{\Psi} (\partial \cdot A + A \cdot\partial)(\slashed{A}\Psi) + h.c.\Big],
\end{equation}
where we have acted with the $\Box$ operators on the respective fields, and assumed that the photon is on-shell, such that $\Box \slashed{A} = -q_{1}^{2}\slashed{A} = 0$. Notice that the sums are identical to 
eq.~(\ref{eq:1stContribution_sum1}). Therefore, writing the hermitian conjugate explicitly, 
eq.~(\ref{eq:2ndContribution1}) reads
\begin{equation}\label{eq:2ndContribution2}
 S_{2} = \frac{ie^{2}}{2} \Bigg( \frac{e^{\frac{k_{2}^{2}}{\Lambda^{2}}}-e^{\frac{k_{1}^{2}}{\Lambda^{2}}}}{k_{2}^{2}-k_{1}^{2}}\Bigg)\int d^{4}x \Big[ \overline{\Psi}(\partial \cdot A + A \cdot \partial) (\slashed{A}\Psi) + (\partial \cdot A + A \cdot \partial)(\overline{\Psi}\slashed{A})\Psi \Big].
\end{equation}
Notice that the second operator acts only on $\overline{\Psi}\slashed{A}$. Acting with the partial derivative on the fermions and the photon will pull down the momentum of the respective field, and one can eliminate the momentum of the photon in favor of the momenta of the two fermions, such that eq. (\ref{eq:2ndContribution2}) becomes
\begin{equation}\label{eq:2ndContribution3}
 S_{2} = \frac{ie^{2}}{2} (k_{1\mu}+k_{2\mu})\Bigg( \frac{e^{\frac{k_{2}^{2}}{\Lambda^{2}}}-e^{\frac{k_{1}^{2}}{\Lambda^{2}}}}{k_{2}^{2}-k_{1}^{2}}\Bigg) \int d^{4}x \overline{\Psi}\Psi A^{\mu}\slashed{A},
\end{equation}
which implies that the Feynman rule corresponding to the second vertex is given by
\begin{equation}\label{eq:2ndContribution4}
    V_{2\mu\nu}(k_{1},k_{2},q_{1},q_{2}) = - e^{2}(k_{1\mu}+k_{2\mu})\gamma_{\nu}\Bigg( \frac{e^{\frac{k_{2}^{2}}{\Lambda^{2}}}-e^{\frac{k_{1}^{2}}{\Lambda^{2}}}}{k_{2}^{2}-k_{1}^{2}}\Bigg).
\end{equation}

Finally, we turn our attention to the last term given in the second line of eq. (\ref{eq:NL_action3}). This part is quite complex, so we resort to some approximations to evaluate it. We first, we notice that
\begin{equation}\label{eq:3rdContribution1}
    (\partial \cdot A + A \cdot \partial) \Box^{l} (\partial \cdot A + A \cdot  \partial) \Box^{n-m-l-2}(\slashed{\partial}\Psi) = ik_{1\nu}\slashed{k}_{2}(q_{1\mu}-k_{2\mu})(-k_{2}^{2})^{n-m-2}A^{\mu}A^{\nu}\Psi,
\end{equation}
where $q_{1\mu}$ is the momentum of one of the photons, and we have assumed that the photons are on-shell and utilized conservation of momentum to eliminate the momenta of the photons in favor of the momenta of the fermions whenever possible. Therefore, the third term in (\ref{eq:NL_action3}) reads
\begin{equation}\label{eq:3rdContribution2}
    S_{3} = \frac{e^{2}}{2}\sum_{n=0}^{\infty}\frac{1}{\Lambda^{2n}n!}\sum_{m=0}^{n-2}\sum_{l=0}^{n-m-2} \int d^{4}x \Big[ (k_{1}^{2m} k_{2}^{2(n-m-2)})k_{1\nu}\slashed{k}_{2}(q_{1\mu}-k_{2\mu})A^{\mu}A^{\nu}\overline{\Psi}\Psi + h.c. \Big].
\end{equation}

We need to evaluate the sums over $l$, $m$ and $n$. First notice that the sum over $l$ is trivial and just lead to a factor of $n-m-2$. Therefore, the sum over $m$ becomes
\begin{eqnarray}\label{eq:3rdContribution_mSum}
    \sum_{m=0}^{n-2}(n-m-2)(k_{1}^{2m}k_{2}^{2(n-m-2)}) = (n-2)\Bigg[ \frac{(k_{2}^{2})^{n-1}-(k_{1}^{2})^{n-1}}{k_{2}^{2}-k_{1}^{2}}\Bigg] \nonumber
    \\
    -\Big(k_{2}^{2(n-2)}\Big)\Big( \frac{k_{1}^{2}}{k_{2}^{2}}\Big) \Bigg[ \frac{1-(n-1)(k_{1}^{2}/k_{2}^{2})^{n-2}+(n-2)(k_{1}^{2}/k_{2}^{2})^{n-1}}{(1-k_{1}^{2}/k_{2}^{2})^{2}}\Bigg],
\end{eqnarray}
and we can now plug this in eq.~(\ref{eq:3rdContribution2}) to evaluate the sum over $n$. The first term in the sum over $n$ yields
\begin{equation}\label{eq:3rdContribution_nSum1}
    \sum_{n=0}^{\infty}\frac{(n-2)}{\Lambda^{2n}n!}\Big( \frac{k_{2}^{2(n-1)}-k_{1}^{2(n-1)}}{k_{2}^{2}-k_{1}^{2}}\Big) = \frac{1}{\Lambda^{2}(k_{2}^{2} - k_{1}^{2})}\Bigg[ \Big(1-\frac{2\Lambda^{2}}{k_{2}^{2}} \Big)e^{\frac{k_{2}^{2}}{\Lambda^{2}}} - \Big(1-\frac{2\Lambda^{2}}{k_{1}^{2}} \Big)e^{\frac{k_{1}^{2}}{\Lambda^{2}}}\Bigg],
\end{equation}
whereas the second term yields
\begin{eqnarray}\label{eq:3rdContribution_nSum2}
    \sum_{n=0}^{\infty}\frac{1}{\Lambda^{2n}n!}\Big(k_{2}^{2(n-2)}\Big)\Big( \frac{k_{1}^{2}}{k_{2}^{2}}\Big) \Bigg[ \frac{1-(n-1)(k_{1}^{2}/k_{2}^{2})^{n-2}+(n-2)(k_{1}^{2}/k_{2}^{2})^{n-1}}{(1-k_{1}^{2}/k_{2}^{2})^{2}}\Bigg] \nonumber \\
    = \frac{1}{(k_{2}^{2}-k_{1}^{2})^{2}} \Bigg[ \Big( \frac{k_{1}^{2}}{k_{2}^{2}}\Big) e^{\frac{k_{2}^{2}}{\Lambda^{2}}} + \Big( \frac{k_{1}^{2}}{\Lambda^{2}} - \frac{k_{2}^{2}}{\Lambda^{2}} + \frac{k_{2}^{2}}{k_{1}^{2}}-2 \Big) e^{\frac{k_{1}^{2}}{\Lambda^{2}}} \Bigg].
\end{eqnarray}

We simplify our results by keeping only the leading order in $\Lambda$, so we drop terms $\sim O(1/\Lambda^{2})$. We plug eqs. (\ref{eq:3rdContribution_nSum1}) and (\ref{eq:3rdContribution_nSum2}) in eq. (\ref{eq:3rdContribution2}) and then evaluate the hermitian conjugate, which can simply be obtained from the first part by interchanging $k_{1} \leftrightarrow k_{2}$. Finally, we arrive at the third contribution to the Feynman rule
\begin{eqnarray}
    \label{eq:3rdContribution3}
    V_{3\mu\nu}(k_{1},k_{2},q_{1},q_{2})  & \simeq & i e^{2}k_{1\nu}\slashed{k}_{2}(q_{1\mu}-k_{2\mu})\Bigg\{ \frac{2}{k_{2}^{2}-k_{1}^{2}} \Bigg( \frac{e^{\frac{k_{1}^{2}}{\Lambda^{2}}}}{k_{1}^{2}} - \frac{e^{\frac{k_{2}^{2}}{\Lambda^{2}}}}{k_{2}^{2}}\Bigg) \nonumber \\
    & + & \frac{1}{(k_{2}^{2}-k_{1}^{2})^{2}} \Bigg[ \Big(\frac{k_{1}^{2}}{k_{2}^{2}}\Big)e^{\frac{k_{2}^{2}}{\Lambda^{2}}} + \Big(\frac{k_{2}^{2}}{k_{1}^{2}}-2\Big)e^{\frac{k_{1}^{2}}{\Lambda^{2}}} \Bigg]\Bigg\} + (k_{1} \leftrightarrow k_{2}).
\end{eqnarray}

Putting all the pieces together from eqs. (\ref{eq:1stContribution2}), (\ref{eq:2ndContribution4}) and (\ref{eq:3rdContribution3}), we arrive at the final result
\begin{flalign}\label{eq:ffAA_Vertex}
&V_{\mu\nu}(k_{1},k_{2},q_{1},q_{2}) \simeq  i e^{2} \Big[(\slashed{k}_{1}+\slashed{k}_{2})g_{\mu\nu} + i (k_{1\mu}+k_{2\mu})\gamma_{\nu} \Big]\Bigg( \frac{e^{\frac{k_{2}^{2}}{\Lambda^{2}}}-e^{\frac{k_{1}^{2}}{\Lambda^{2}}}}{k_{2}^{2}-k_{1}^{2}}\Bigg) \nonumber \\
    & +   k_{1\nu}\slashed{k}_{2}(q_{1\mu}-k_{2\mu})\Bigg\{ \frac{2}{k_{2}^{2}-k_{1}^{2}} \Bigg( \frac{e^{\frac{k_{1}^{2}}{\Lambda^{2}}}}{k_{1}^{2}} - \frac{e^{\frac{k_{2}^{2}}{\Lambda^{2}}}}{k_{2}^{2}}\Bigg) + \frac{1}{(k_{2}^{2}-k_{1}^{2})^{2}} \Bigg[ \Big(\frac{k_{1}^{2}}{k_{2}^{2}}\Big)e^{\frac{k_{2}^{2}}{\Lambda^{2}}} + \Big(\frac{k_{2}^{2}}{k_{1}^{2}}-2\Big)e^{\frac{k_{1}^{2}}{\Lambda^{2}}} \Bigg]\Bigg\} \nonumber \\
    & +   k_{2\nu}\slashed{k}_{1}(q_{1\mu}-k_{1\mu})\Bigg\{ \frac{2}{k_{1}^{2}-k_{2}^{2}} \Bigg( \frac{e^{\frac{k_{2}^{2}}{\Lambda^{2}}}}{k_{2}^{2}} - \frac{e^{\frac{k_{1}^{2}}{\Lambda^{2}}}}{k_{1}^{2}}\Bigg) + \frac{1}{(k_{1}^{2}-k_{2}^{2})^{2}} \Bigg[ \Big(\frac{k_{2}^{2}}{k_{1}^{2}}\Big)e^{\frac{k_{1}^{2}}{\Lambda^{2}}} + \Big(\frac{k_{1}^{2}}{k_{2}^{2}}-2\Big)e^{\frac{k_{2}^{2}}{\Lambda^{2}}} \Bigg]\Bigg\}.    
\end{flalign}

\end{document}